%% file: main.tex
\documentclass[10pt,twocolumn,letterpaper]{article}

\usepackage[pagenumbers]{cvpr} 

\input{preamble}

\definecolor{cvprblue}{rgb}{0.21,0.49,0.74}
\usepackage[pagebackref,breaklinks,colorlinks,citecolor=cvprblue]{hyperref}

\def\paperID{216} 
\def\confName{3DV\xspace}
\def\confYear{2026\xspace}

\title{Layered Quantum Architecture Search for 3D Point Cloud Classification}

\author{Natacha Kuete Meli$^1$
\hspace{1cm}
Jovita Lukasik$^1$
\hspace{1cm}
Vladislav Golyanik$^2$
\hspace{1cm}
Michael Moeller$^1$\\
$^1$University of Siegen 
\hspace{1cm}
$^2$MPI for Informatics, SIC\\
{\tt\small vsa.informatik.uni-siegen.de}
\hspace{1cm}
{\tt\small 4dqv.mpi-inf.mpg.de}
}

\begin{document}
\maketitle
\input{0_abstract}    
\input{1_intro}
\input{2_related_work}

\input{3_method}
\input{4_results}
\input{5_summary}
{
    \small
    \bibliographystyle{ieeenat_fullname}
    \bibliography{main}
}
\input{X_suppl}
\end{document}

%% file: preamble.tex
\usepackage[dvipsnames]{xcolor}

\usepackage{graphicx}
\usepackage{tikz}
\usepackage{quantikz}
\usetikzlibrary{positioning,patterns,3d,angles,quotes}
\usepackage{pgfplots}
\pgfplotsset{compat=1.18}
\usepackage{upgreek}
\usepackage{ifthen}
\usepackage{calc}
\usepackage{amsmath,amssymb,amsthm,times, bm}
\usepackage{bbm}
\usepackage{graphicx}
\usepackage{multicol} 
\usepackage{multirow}
\usepackage{caption}
\usepackage{subcaption}
\usepackage{braket}
\usepackage{pifont}
\usepackage{fontawesome}
\usepackage{xifthen} 

\usepackage{algorithm}
\usepackage[noend]{algpseudocode}

\usepackage{wrapfig}

\newcommand{\N}{\mathbb{N}}

\newcommand{\R}{\mathbb{R}}

\newcommand{\mat}[1]{\mathbf{#1}}
\newcommand{\mcal}[1]{\mathcal{#1}}

\newcommand{\CRX}[0]{\mathrm{\textsc{CRX}}}
\newcommand{\RX}[0]{\mathrm{\textsc{RX}}}
\newcommand{\RY}[0]{\mathrm{\textsc{RY}}}
\newcommand{\RZ}[0]{\mathrm{\textsc{RZ}}}

\DeclareMathOperator*{\argmax}{arg\,max}
\newcommand{\Matrix}[2][0.8]{%
	\setlength{\arraycolsep}{5pt}\renewcommand{\arraystretch}{#1}
	\left(\hskip-5pt\begin{array}{*{10}{c}}#2\end{array}\hskip-5pt\right)}

%% file: 0_abstract.tex
\begin{abstract}
We introduce layered Quantum Architecture Search (layered-QAS), a strategy inspired by classical network morphism that designs Parametrised Quantum Circuit (PQC) architectures by progressively growing and adapting them. 
PQCs offer strong expressiveness with relatively few parameters, yet they lack standard architectural layers (e.g., convolution, attention) that encode inductive biases for a given learning task. 
To assess the effectiveness of our method, we focus on 3D point cloud classification as a challenging yet highly structured problem. 
Whereas prior work on this task has used PQCs only as feature extractors for classical classifiers, our approach uses the PQC as the main building block of the classification model. 
Simulations show that our layered-QAS mitigates barren plateau, outperforms quantum-adapted local and evolutionary QAS baselines, and achieves state-of-the-art results among PQC-based methods on the ModelNet dataset
\footnote{project page: \url{https://4dqv.mpi-inf.mpg.de/LQAS/}}.
\end{abstract}

%% file: 1_intro.tex
\section{Introduction}
\label{sec:intro}

Quantum Machine Learning (QML)~\cite{biamonte2017quantum,cerezo2022challenges,kuete2025quantum}, through shallow-depth Parametrised Quantum Circuits (PQCs), is anticipated to pave the way for utility-scale quantum computing.
PQCs operate in high-dimensional Hilbert spaces with comparably few parameters, leveraging superposition and entanglement to extract features beyond classical neural networks.
However, designing effective PQC architectures for task-oriented feature extraction remains challenging.
While classical models benefit from well-established architectural layers and inductive biases, PQCs lack such standardised building blocks.
In addition, the same quantum properties that enable high expressiveness can also cause barren plateaus~\cite{larocca2024review}, where flattened loss landscapes hinder optimisation.
Furthermore, PQCs are largely composed of linear transformations, missing the non-linear activations that underpin classical deep networks’ representational power.
In light of this, maximising PQC architectural design is crucial to ensure discriminative feature extraction while keeping the PQCs trainable and robust.

\begin{figure}
    \centering
    \includegraphics[width=1.08\linewidth, trim={0.13cm .05cm 0cm -.05cm}, clip]{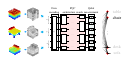}
    \caption{
    \textbf{Overview of our framework for 3D point classification using Parametrised Quantum Circuits (PQCs).} 
    The input point cloud is voxelised and used to prepare the quantum system.
    We then use the new layered Quantum Architecture Search (layered-QAS) approach to engineer the PQC design that meaningfully learns features from the encoded point cloud. 
    Lastly, qubits are measured to extract the learned features, which are used by an optionally learnable classical linear layer for classification.
    }
    \label{fig:teaser}
\end{figure}

Quantum Architecture Search (QAS) offers a systematic way to address these design challenges by automatically exploring the space of PQC structures~\cite{zhu2022brief,lu2023qas,martyniuk2024quantum}.  
By replacing manual trial-and-error with guided search, QAS can discover expressive circuits for task-specific feature extraction, improve trainability, and mitigate barren plateaus.  
Existing QAS policies, however, face several limitations. 
Super-circuit and weight-sharing strategies~\cite{du2022quantum,wang2022quantumnas} not only search in very large spaces that are computationally demanding, but also often converge to suboptimal architectures, as candidate models can adversarially update the super-circuit parameters. 
Evolutionary and RL-based methods may struggle with scalability or training instability due to sensitivity to hyperparameters and reward design~\cite{ostaszewski2021reinforcement}, while differentiable approaches~\cite{wu2023quantumdarts} are prone to getting trapped in local optima. 
Moreover, none of these QAS methods account for PQC-specific considerations such as strategic training~\cite{skolik2021layerwise,duffy2024quantum} and careful initialisation~\cite{grant2019initialization,kashif2024dilemma,wang2024trainability}, which have been shown to mitigate barren plateaus, improve generalisation, and enhance feature learning.

In response to these limitations, we introduce \emph{layered-QAS}. Building on ideas from the Lamarckian-based LEMONADE~\cite{elsken2018efficient}, our policy adds new layers to a pre‑trained circuit and retains only those that provide the largest performance improvement. 
To further increase efficiency, we incorporate a pruning mechanism that removes gates operating near the identity, reducing unnecessary complexity without sacrificing expressivity. 
Together, these components make layered‑QAS a robust framework for discovering high‑performing PQCs.

We target 3D point cloud classification as a demanding yet structured exemplary application to evaluate our approach.
The task involves assigning labels to objects represented by unordered sets of points, with applications in autonomous driving, robotics, and semantic segmentation~\cite{zhang2023deep,sarker2024comprehensive,muzahid2024deep}.
Point clouds vary in density, resolution, and shape, while modern sensors produce increasingly large datasets.
These characteristics make point cloud classification an ideal benchmark for QML, as it requires models to reason about spatial relationships in irregular, high-dimensional data.
In this context, sQCNN-3D~\cite{baek2023stereoscopic} applies quanvolutional filters to point cloud patches but delegates classification to a classical fully connected network—leaving the quantum model underutilised.
In contrast, we pursue a fully quantum classification pipeline, minimising classical post-processing and improving feature extraction by the quantum circuit through architecture search.
Applied to 3D multi-class point cloud classification, our pipeline enables near end-to-end quantum classification with only minimal classical components.
An overview of our 3D classification framework is provided in~\Cref{fig:teaser}.
To summarise, our main contributions are as follows: 
\begin{itemize}
    \item A new layered-QAS policy for discovering improved and task-adapted PQC designs, avoiding manual trial-and-error approaches (Sec.~\ref{sec:method}; Sec.~\ref{ssec:LQAS}); 
    \item A new framework for 3D point cloud classification based on PQCs and amplitude encoding of 3D data, leveraging layered-QAS for architectural design (Sec.~\ref{sec:workflow}). 
\end{itemize} 
We experiment on the ModelNet datasets~\cite{wu20153d} using a quantum computer simulator and show that our models outperform the existing sQCNN-3D quantum baseline~\cite{baek2023stereoscopic} and are even competitive with a classical baseline of similar expressivity that we design. 
Our search strategies can find parameter-efficient PQC architectures. 
Moreover, our layered search achieves better performance than the evolutionary search and a QAS-adapted local search~\cite{white2021exploring}, a baseline that incrementally improves PQC architectures by making small localised changes to the PQC and retains the changes that enhance performance. 

%% file: 2_related_work.tex
\section{Related Work}
\label{sec:related_work}

\subsection{PQC Training and QAS}

\paragraph{Strategic PQC Training.}
Strategic training of PQCs is believed to dampen trainability issues. 
Skolik et al.~\cite{skolik2021layerwise} proposed a layer-wise training approach, which incrementally grows the circuit depth, freezes previously learned parameters, and optimises only the parameters of added layers.
This layered training strategy not only speeds up the training but also experimentally proved to increase the generalisation error.
Similarly, the method proposed by Duffy et al.~\cite{duffy2024quantum} allows not only for the addition of new parametrised gates, but also feature-map encodings of the data to incrementally grow the circuits.
Data re-uploading is proven to increase the expressivity of PQCs~\cite{schuld2021effect}.
Both methods suggest that incrementally training PQCs is a promising approach to enhance PQCs' expressivity without compromising their trainability.
Note that a layered training approach was also used by Krahn et al.~\cite{krahn2024qpsgbd} for binary networks with quantum annealed gradients.

\paragraph{Strategic PQC Initialisation.}
Random initialisation of PQCs was shown to be one cause of barren plateaus~\cite{mcclean2018barren,haug2021capacity}.
Grant et al.~\cite{grant2019initialization} proposed a selective initialisation method that randomly selects only a subset of the initial parameter values and chooses the remaining ones so that the circuit is a sequence of shallow blocks that evaluate to the identity, limiting the circuit depth in the first parameter update.
Wang et al.~\cite{wang2024trainability} proved that reducing the initial domain of each parameter inversely proportional to the square root of the circuit depth causes the magnitude of the cost gradient to decay at most polynomially with respect to the qubit count and the circuit depth.
Kashif et al.~\cite{kashif2024dilemma} empirically showed that initialising parameters within smaller distribution ranges, with lower magnitudes, tends to perform better than using larger ranges with higher magnitudes.

\paragraph{Quantum Architecture Search (QAS).}
QAS methods often use a super-circuit to define the search pool of candidate circuits, as seen in early works by Du et al.~\cite{du2022quantum} and Wang et al.~\cite{wang2022quantumnas}, where training updates the super-circuit before selecting and fine-tuning the best candidate. Ma et al.~\cite{ma2024continuous} enhanced this with an evolutionary post-training process, while other evolutionary approaches forgo super-circuits, training architectures from scratch~\cite{zhang2023evolutionary,chivilikhin2020mog}. 
Weight-sharing improves memory efficiency but may cause suboptimal convergence. Differentiable QAS methods, like those by Wu et al.~\cite{wu2023quantumdarts} and Zhang et al.~\cite{zhang2022differentiable}, optimize the search domain, while Reinforcement Learning (RL)-based methods~\cite{ostaszewski2021reinforcement,ye2021quantum,kuo2021quantum,dai2024quantum,kundu2024kanqas} use neural agents to identify effective architectures.
However, differentiable methods risk favoring local optima, and RL methods may face instability due to hyper-parameters and reward function designs.

Our LEMONADE-\cite{elsken2018efficient}-inspired \emph{layered-QAS} combines informed initialisation with progressive layer-wise training. (i) Candidate layer architectures are briefly trained and ranked to discard weak options; 
(ii) when new layers are added, previously learned parameters remain trainable and continue to improve. 
This warm-starting lets child models inherit and refine the performance of their parents.

\subsection{Classification using PQCs} 

\paragraph{PQC Classification and Architectures.} Few methods have explored 3D point cloud classification with PQCs, as most prior QML works focused on binary classification on small datasets like Moons, Iris or heavily downsampled MNIST datasets; see Refs.~\cite{fan2023hybrid,bowles2024better,kuete2025quantum} for an overview.

Binary and 2D classification have long served as standard PQC benchmarks. 
Farhi \etal~\cite{farhi2018classification} and Cong \etal~\cite{cong2019quantum} were among the first to design quantum neural networks for binary classification. 
Havlíček \etal~\cite{havlivcek2019supervised} proposed a quantum variational classifier using a feature-map encoding to project data into high-dimensional spaces.
Henderson \etal~\cite{henderson2020quanvolutional} introduced quanvolutional filters for PQC-based patch feature extraction. 
Salinas \etal~\cite{perez2020data} developed a universal classifier using data re-uploading, showing that a single-qubit PQC can serve as a universal classifier when target classes correspond to specific Bloch-sphere states—though in multi-class settings, class-state non-orthogonality induces correlations. 
More recent works~\cite{jing2022rgb,kuros2022traffic,senokosov2024quantum} extended these models to image datasets such as MNIST~\cite{deng2012mnist}, CIFAR~\cite{krizhevsky2009learning}, and GTSRB~\cite{Houben-IJCNN-2013}.
To the best of our knowledge, the work by Baek et al.~\cite{baek2023stereoscopic} is the only PQC approach to 3D point cloud classification. 
Their method, sQCNN-3D, employs multiple PQCs that are trained in parallel and behave as filters for feature extraction.
Those so-called quanvolution filters process small patches of the voxelised cloud encoded using angle encoding a data re-uploading. 
To mitigate barren plateaus, these PQCs remain small in size and depth. 
Similar to classical convolutional filters, each PQC uses the same parameters to process all the inputted patches. 
Features are extracted by locally measuring the qubits and concatenated into a feature vector that is further processed by an MLP, which has significantly more parameters than the PQCs. 
Due to this overwhelming classical part of the model, it is unclear whether the PQCs still play a role in the model's performance.

In contrast, we propose a quantum-driven alternative that minimises classical components: voxelised inputs are amplitude-encoded into PQCs, features are extracted directly via quantum measurements, and classification is performed with at most one final classical linear layer.

%% file: 3_method.tex
\section{3D Point Classification with Layered-QAS}
\label{sec:method}

This section presents our layered-QAS policy for PQC design. 
To set the stage, we first outline the classification pipeline in~\Cref{sec:workflow}, followed by the presentation of our layered-QAS policy in~\Cref{sec:layered}. 
A brief overview of gate-based quantum computing is provided in~\Cref{app:intro_qc}.

\subsection{Workflow}
\label{sec:workflow}

\subsubsection{3D Point Cloud Encoding}
\label{sec:encoding}
\begin{figure}
    \centering
	\includegraphics[scale=.37, trim={.5cm .05cm 0cm .05cm}, clip]{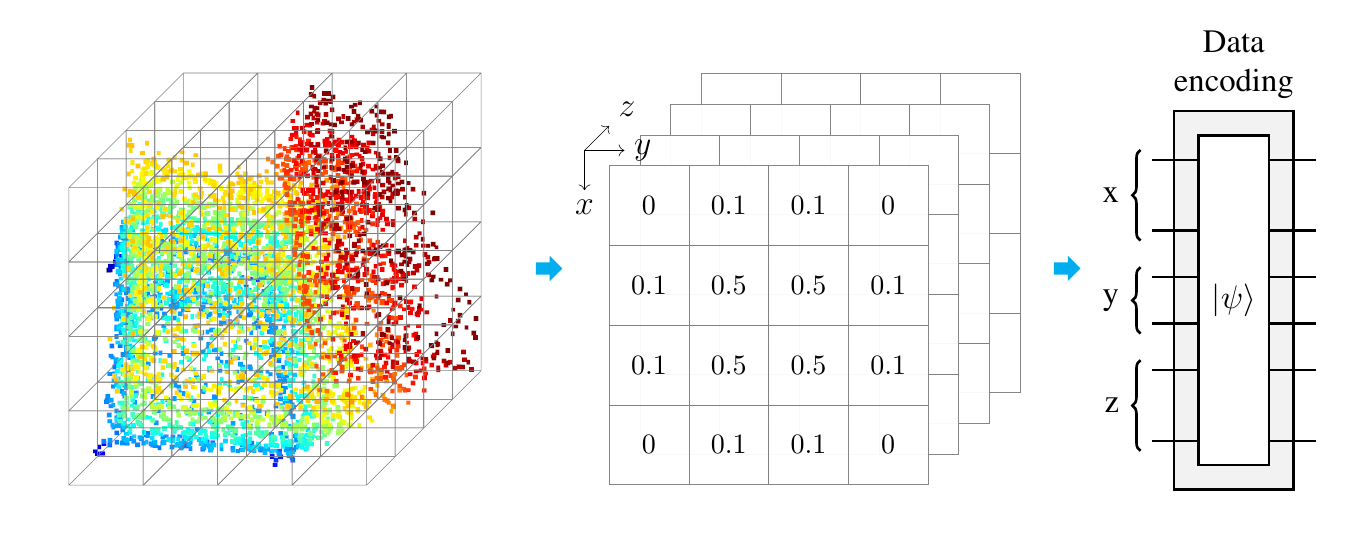}
    \caption{\textbf{Amplitude encoding of a 3D point cloud into a quantum state.}
    We partition the point cloud into voxels (left) and compute the normalised density of each voxel, \ie the proportion of points within the voxel (middle).
    This creates a matrix that we vectorise and use as input state vector for the PQC (right).
    The granularity of the voxelisation, which defines the number of qubits used, is user-defined.
    Colours are added for visualisation purposes.
    }
    \label{fig:data_encoding}
\end{figure}

We use amplitude encoding as described in~\Cref{fig:data_encoding} to encode the input point cloud in a quantum state vector.

Given is a point cloud $\tilde{\mcal P} = \{\tilde p^i = (\tilde p_x^i, \tilde p_y^i, \tilde p_z^i)\}_{i=1}^N$ and a voxel granularity $k\in \N$.
First, we normalise the point cloud $\tilde{\mcal P}$ into $\mcal P = \{p^i = (p_x^i, p_y^i, p_z^i)\}_{i=1}^N$ by fitting it into the 3D unit cube that we later scale by $2^k-1$.
We then partition this cube in each dimension into $2^k$ sub-cubes called voxels. 
Next, for each voxel that we index by integer coordinates $(x, y, z)$, we compute the proportion or density
\begin{equation}
\delta_{xyz} = \frac{1}{N} \sum_{i=1}^{N} \mathbb{I}\big( p_x^i \in V_x, p_y^i \in V_y, p_z^i \in V_z \big),
\end{equation}
of points within it, with $V_x = [x, x+1)$, $V_y = [y, y+1)$, and $V_z = [z, z+1)$ representing the voxel boundaries, and $\mathbb{I}(\cdot)$ returning $1$ if the input condition is true, and $0$ otherwise.  
Finally, we use the computed densities to form a normalised state vector 
\begin{equation}
\label{eq:state_vector}
    \ket{\psi} = \sum_{x, y, z} \sqrt{\delta_{xyz}} \ket{xyz}
\end{equation}
in which we prepare the quantum system before the PQC transformation, with $\ket{xyz} = \ket{x} \otimes \ket{y} \otimes \ket{z}$ being a composed system of three registers of $k$ qubits each that encode the $x, y$ and $z$ coordinates of the voxels in the binary basis. 
Thus, the overall encoding consists of $3k$-many qubits.
Since the point cloud occupies only a subregion of the cube, the state vector in~\Cref{eq:state_vector} is sparse within the $2^{3k}$-dimensional Hilbert space, enabling a relatively efficient state-preparation~\cite{schuld2018supervised,Li2024NearlyOC}.
Example voxelised point clouds for different $k$ are visualised in~\Cref{app:voxelisation}.

\subsubsection{PQC Transformation}  
\label{sec:transformations}
The encoded quantum state vector undergoes several layers of parametrised quantum gates forming the PQC, see~\Cref{fig:teaser}.
These gates transform the initial state vector into a final state that can be measured and post-processed.
A PQC of $\ell$ layers can be densely represented as an operator $\mat U(\bm \theta)$, with $\bm \theta$ being the set of parameters to be optimised.
It transforms the initial state vector into a parametrised state
\begin{equation} 
\label{eq:pqc_trans}
\ket{\psi(\bm \theta)} = \mat U(\bm \theta)\ket{\psi} = \mat L_\ell(\theta_\ell) \cdots \mat L_1(\theta_1) \ket{\psi}.
\end{equation}
Each layer $\mat L_i(\theta_i)$ consists of a combination of single-qubit and/or multi-qubit gates. 
The single-qubit gates we use are Pauli $\RX, \RY, \RZ$ rotations:
\begin{align} 
\RX(2\theta) &= \Matrix{\cos(\theta) & -i\sin(\theta) \\ -i\sin(\theta) & \cos(\theta)}, \\
\RY(2\theta) &= \Matrix{\cos(\theta) & -\sin(\theta) \\ \sin(\theta) & \cos(\theta)}, \quad \text{and}\\
\RZ(2\theta) &= \Matrix{e^{-i \theta} & 0 \\ 0 & e^{i \theta}}. 
\end{align}
Multi-qubit entangling gates introduce correlations between qubits.
We use two-qubit controlled $\RX$ rotations
\begin{equation} 
\textsc{CRX}(2\theta) = \Matrix{
1 & 0 & 0 & 0 \\ 
0 & 1 & 0 & 0 \\ 
0 & 0 & \cos(\theta) & -i\sin(\theta) \\ 
0 & 0 & -i\sin(\theta) & \cos(\theta) 
} .
\end{equation}
This set of gates is similar to the well-known universal set $\{\RX, \RY, \RZ, \text{Phase}, \textsc{CNOT}\}$~\cite{williams2011quantum,nielsen2010quantum}.
The optimal parameters $\bm \theta$ for the classification task should be found by minimising the classification loss, while the circuit architecture will be designed by automated layered-QAS.

\subsubsection{Measurement}  
We perform local qubit measurements by measuring each qubit individually.  
As the PQC uses the three Pauli $\RX, \RY, \RZ$ rotations to position the qubits on the Bloch sphere, we measure in the Pauli $\textsc{X}, \textsc{Y}, \textsc{Z}$ bases.  

The Pauli observables $\textsc{M}_q$, with \mbox{$\textsc{M} \in \left\{\textsc{X}, \textsc{Y}, \textsc{Z}\right\}$} applied on the $q$th qubit with the identity to the remaining qubits, measure the probability distribution for outcomes in the $\textsc{M}$ bases. 
The corresponding expectation values 
\begin{equation}
    \braket{\textsc{M}_q(\bm \theta)} = \braket{\psi(\bm \theta)|\textsc{M}_q|\psi(\bm  \theta)}
\end{equation}
are returned for the three bases, with $\ket{\psi(\bm \theta)}$ from~\Cref{eq:pqc_trans} being the state of the system after the PQC transformation.
In total, the expectation values returned constitute a set of $3 \cdot 3k$ learned features, with $k$ being the number of qubits of each coordinate register.

\subsubsection{The Loss Function} 
Let $c$ be the number of class labels.  
As mentioned above, the QPC outputs $3 \cdot 3k$ learned features, which may not correspond to the number $c$ of classes.  
We use a classical linear layer to map these features into a length-$c$ logit vector, which is then compared by the cross-entropy loss function 
\begin{equation}
    \mathcal{L}_{\text{CE}} = - \sum_{j=1}^{c} y_j \log \hat{p}_j
\end{equation}
to the one-hot encoded ground-truth label $\mathbf{y} = (y_1, \dots, y_c)$, with $\hat{\mathbf{p}} = (\hat{p}_1, \dots, \hat{p}_c)$ being the predicted probability distribution obtained by applying softmax to the logits.

Ideally, the linear layer is trainable so it can learn weights that map PQC outputs to the correct logits. 
To assess whether the PQCs themselves learn meaningful features, we also test a non‑trainable linear layer implemented as a fixed random Gaussian projection, i.e., a randomly initialized matrix with entries drawn from a normal distribution.

\begin{algorithm}[t]
\caption{Layered Quantum Architecture Search}\label{alg:algorithm_layered}
\begin{algorithmic}[1]

\Require Layer architectures, number $T$ of layer types, training and validation sets, ranking metric $\text{Best}$.

\State $ \mat U_0 := \mat I$

\For{$i = 0, 1, 2, \ldots$} 
    \State $\text{ArchList} := [\ ]$
    \Comment{Empty list}
    \State Layer type $t := i \mod T$
    \State $\mcal L_t  = \{\text{Random set of layers of type t}\}$
    \For{$\mat L_{i+1} \in \mcal L_t$} 
        \State $\mat U_{\text{candidate}} = \mat L_{i+1} \mat U_i$
        \State $\text{Train} \ \mat U_{\text{candidate}} \ \text{for a few epochs}$
        \State Append $\mat U_{\text{candidate}}$ to ArchList
    \EndFor
    \State Update $ \mat U_{i+1} = \text{Best} (\text{ArchList})$
    \Comment{Ranking}
\EndFor
    \State \textbf{Return} $ \mat U_{i+1}$
\end{algorithmic}
\end{algorithm}

\subsection{Layered Quantum Architecture Search}\label{ssec:LQAS}

\label{sec:layered}
\begin{figure}
    \centering
    \includegraphics[width=0.99\linewidth]{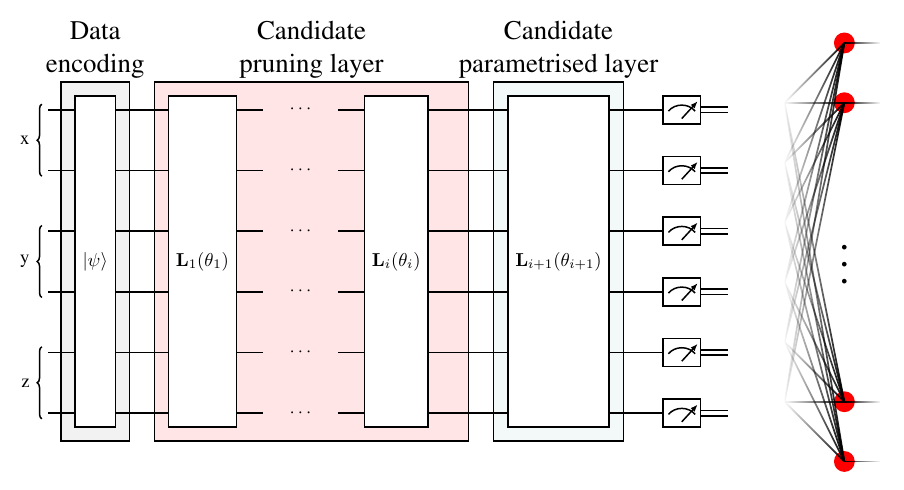}
    \caption{\textbf{Workflow of our layered-QAS.}
    At generation $(i+1)$, the algorithm expands the PQC with a layer $\mat L_{i+1}(\theta_{i+1})$, chosen from single‑qubit, entangling, or pruning layers. 
    Single‑qubit and entangling layers add new parametrised gates (green), while pruning layers remove gates (red). 
    Several candidate layers are evaluated, and the most effective circuit extension is retained.
    }
    \label{fig:workflow}
\end{figure}

We propose to find a suitable combination of the gates described in~\Cref{sec:transformations} for a PQC architecture via a new layered search methodology that is outlined in~\Cref{alg:algorithm_layered} and illustrated in~\Cref{fig:workflow}. 
We denote by $\mat {U}_i(\bm \theta_i)$ the unitary operator of the PQC at generation $i$.  

We begin with a trivial identity circuit $\mat {U}_0 (\bm \theta_0) = \mat {I}$, which consists of data encoding followed by measurement.  
At each generation $(i+1)$, the existing circuit $\mat {U}_i(\bm \theta_i)$ is expanded by adding a new layer $\mat {L}_{i+1}(\theta_{i+1})$.  
To explore the best possible architecture, different designs of the same layer type are tested, each yielding a new PQC candidate:  
\begin{equation}
    \mat {U}_{\text{candidate}}(\bm \theta_{i+1}) = \mat {L}_{i+1}(\theta_{i+1}) \mat {U}_i(\bm \theta_i).
\end{equation}
The type of layer alternates across generations.
For a fair comparison, the parameters of the previous generation remain unchanged in all candidate PQCs.  
Furthermore, newly added layers are designed to almost preserve the accuracy of $\mat {U}_i(\bm \theta_i)$ at the start of the training.  
This is achieved, for example, by initialising the parameters of added layers to zero, ensuring that they initially act as the identity.  

Each PQC candidate $\mat {U}_{\text{candidate}}(\bm \theta_{i+1})$ undergoes a few training epochs.  
After training, candidates are ranked based on their highest performance on the validation set during the last training epoch.
The most effective circuit  
\begin{equation}
    \mat {U}(\bm \theta_{i+1})= \argmax_{\mat {U} \in \{\mat {U}_{\text{candidate}}\}} \ \text{Best}(\mat {U}_{\text{candidate}}(\bm \theta_{i+1})),
\end{equation}
along with its optimised parameters $\bm \theta_{i+1}$, is selected for the next generation. 
Our ranking metric $\text{Best}(\cdot)$ is the classification accuracy on the validation data.
The process iterates, progressively building a more and more expressive and effective PQC by systematically selecting the best-performing circuits from each generation.

\paragraph{Layer Architectures.} 
The layered architectures we consider for the point cloud classification are illustrated in~\Cref{fig:candidate_layers}.  
Three types of layers are considered, i.e.,~single-qubit layers, entangling layers, and pruning layers:

\begin{figure}
    \centering
	\includegraphics[scale=.73, trim={.1cm 0cm .1cm 0cm}, clip]{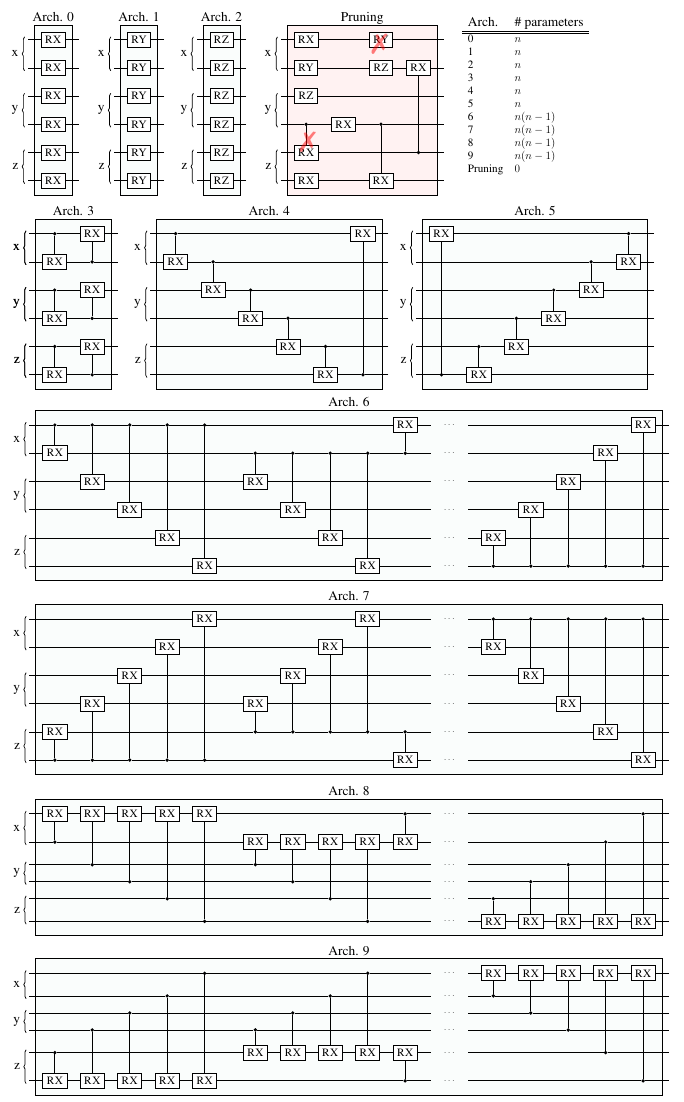}
    \caption{\textbf{Layers tested in our layered-QAS.} 
    The number of parameters for each layer is shown at the top right. 
    Special architectures include the pruning layer, which deletes gates, and architecture 3, which circularly applies CNOTs on coordinate registers. 
    } 
    \label{fig:candidate_layers} 
\end{figure} 

\begin{itemize}
    \item \textbf{Single-qubit layers} 
    apply single-qubit Pauli $\RX, \RY, \RZ$ rotation gates, enabling the circuit to learn single-qubit transformations.  
    Candidate layers at a given generation are architectures $0$, $1$ and $2$.

    \item \textbf{Entangling layers}  introduce controlled rotations between qubit pairs, allowing the circuit to learn entanglement patterns and enhance its expressivity~\cite{sim2019expressibility}.
    We experiment with $\CRX$ gates. 
    Although some architecture pairs in \Cref{fig:candidate_layers} (e.g., $4\&5$, $6\&7$, $8\&9$) appear similar, they differ in how control is applied. 
    In architecture $4$, each qubit except the first is controlled by all its predecessors, whereas in architecture $5$ it is controlled only by its immediate predecessor. 
    At each generation, three entangling-layer candidates are sampled uniformly at random from architectures $3$ to $9$.

    \item \textbf{Pruning layers} 
    randomly remove a proportion of variational gates from the previous circuit $\mat{U}_i(\bm{\theta}_i)$ if the absolute value of their parameters falls below a predefined dropout threshold. 
    The intuition is that gates with small rotation angles contribute minimally to the computation while increasing circuit complexity.  
    By eliminating them, we reduce computational cost without significantly sacrificing the accuracy.  
    Candidate pruning layers at a given generation are three pruning layers that randomly select gates to prune among the ones with small angles.
\end{itemize}
Layer types are explored cyclically. 
We, however, note that different layer types could also go into the candidate dimension. 
Our layered search enables a structured and principled exploration of the PQC architectures, progressively refining their performance while maintaining efficiency. 

%% file: 4_results.tex
\section{Experimental Results}
\label{sec:results}
In this section, we present the results of our layered-QAS strategy in the context of 3D point cloud classification. 
After presenting the implementation and evaluation details, we begin in~\Cref{sec:benchmark}, by summarising the 3D classification results of the layered-QAS to evaluate the performance of our quantum model in comparison with quantum and classical baselines. 
Next, in~\Cref{sec:qas_benchmark}, we benchmark our layered search against alternative QAS policies. Finally, in~\Cref{sec:ablation}, we justify key design choices through ablation studies on the voxel granularity and the gate pruning threshold, and briefly discuss runtimes.

\paragraph{Implementation Details.} 
The code is written in Python using the PennyLane framework~\cite{bergholm2018pennylane}. 
In PennyLane, the quantum tape ---sequence of gates in a quantum circuit--- is stored in a Python list that can be easily modified to alter the circuit architecture.
All experiments are simulated and performed in the idealised noise-free setting, and gradients for training are computed with automatic differentiation. 
We use a NVIDIA GeForce RTX 4090 GPU.

\paragraph{Evaluation Methodology.} 
We experiment on the ModelNet10 and ModelNet40 datasets~\cite{wu20153d}, with $10$ and $40$ labels respectively, to classify the point clouds into. 
ModelNet samples are object triangle meshes, from which we uniformly sample 5000 3D points at random to generate object point clouds for classification. 
ModelNet10 originally has $3991$ training and $908$ test samples, while ModelNet40 has $9842$ training and $2468$ test samples. 
We take out $20\%$ of the original training samples to form the validation set on each dataset and keep the test set intact for evaluation.
The granularity of the voxelisation is $k=3$ in each of the $x, y$ and $z$ dimensions, leading to $9$ qubits on which the PQCs operate.
We reduce the datasets by $90\%$ during the architecture search phase, but use the full datasets for fine-tuning after the search.
We observed that the search on full, large datasets causes the parameters in the evolutionary search in~\Cref{sec:qas_benchmark} to deeply adapt to the candidate models, resulting in highly conflicting updates to the super-circuit parameters (see the discussion in~\Cref{app:modelnet10_fullds}). 
Conversely, the search on smaller datasets enables the selection of candidate models that generalise better. 
Benchmark classification results are reported based on the full datasets.

All models are optimised with the ADAM optimiser~\cite{kingma2014adam} and a learning rate $\text{lr}=0.1$ in the search phase and $\text{lr}=0.03$ in the fine-tuning phase.
Each candidate model is trained for $5$ epochs.
The classification performance metric is the standard top-1 accuracy.
The gate pruning threshold is $\pi/10$ in our layered-QAS.

\subsection{Resuts on ModelNet10\&40}
\label{sec:benchmark}
We benchmark our approach in a shallow version obtained after 10-layer search iterations, as well as a deeper version after 20 search iterations on the full ModelNet10 and ModelNet40 datasets against the following competitors:
\begin{itemize}
    \item The only prior work on PQCs for 3D point cloud classification, the sQCNN-3D method~\cite{baek2023stereoscopic}, with 2 quanvolution filters as tested in the original publication.
    \item A vanilla CNN with one $3$D convolution, one ReLU layer and one fully-connected layer to keep the parameter count and expressivity of the models comparable.
    For instance, the 3D convolutional layer is a linear transformation as the PQCs, the ReLU activation is non-linear as the measurement of the quantum systems, and the last linear layer is used for classification.
\end{itemize}
We train the benchmark models for $20$ epochs with a higher learning rate on the reduced dataset, and subsequently fine-tune all models over $10$ more epochs with a lower learning rate on the full datasets.

As we can see from the results in~\Cref{tab:benchmark}, both variants of the proposed approach outperform the prior work on 3D point cloud classification with PQCs as well as the simple CNN baseline on both benchmark datasets in terms of validation and test accuracy. Remarkably, our reduced PQC model after 10 layered search iterations obtains high accuracy at less than half the number of learnable quantum parameters, trailing the deeper (20-iteration) model but surprisingly few percents only. 
To get insight into shapes that are challenging to classify, we show confusion matrices on ModelNet10 in~\Cref{app:confusion_modelnet10}.

\begin{table*}[th]
\centering
\begin{tabular}{lcccccccccc}
                \hline
                \hline
                & \multicolumn{5}{l}{\textbf{ModelNet10}}& \multicolumn{5}{l}{\textbf{ModelNet40}} \\
                & \textsc{NPQ}  & \textsc{NPC}   & \textsc{Tr.}     & \textsc{Va.}  & \textsc{Ts.}
                & \textsc{NPQ}  & \textsc{NPC}   & \textsc{Tr.}     & \textsc{Va.}  & \textsc{Ts.}\\
                \hline
Layered$_{10}$ [Ours]          &   $100$    & $270$    &    $\mathbf{92\%}$      &   $\underline{92}\%$      &    $\underline{84}\%$               &   $105$    & $1080$    &    $59\%$       &   $\underline{56}\%$    &   $\underline{54}\%$               \\
Layered$_{20}$ [Ours]          &   $279$    & $270$    &    $\mathbf{92\%}$      &   $\mathbf{93\%}$      &    $\mathbf{85\%}$               &   $279$    & $1080$    &    $\mathbf{61}\%$       &   $\mathbf{59}\%$    &   $\mathbf{55} \%$                        \\ 
sQCNN-3D ~\cite{baek2023stereoscopic}& $48$   &    $650$     &     $83 \%$      &    $79 \%$           &  $72 \%$  & $48$   &    $2600$     &     $47 \%$      &    $45 \%$           &  $41 \%$  \\ 
Vanilla CNN          &   $0$    & $590$    &    ${90} \%$      &   $88 \%$      &    $82 \%$               &   $0$    & $1580$    &    $\underline{60} \%$      &   $55 \%$      &    $54 \%$ \\  
 \hline   
\end{tabular}
    \caption{\textbf{Resources and performance comparison for the models.}
    Keys: \textsc{``NP(Q/C)''}=Number of parameters in the quantum or classical backbones; 
    \textsc{``Tr./Va./Ts.''}=Train/Validation/Test top-1 accuracy;
    The subscript in our layered approach refers to the number of layered-QAS iterations. Our proposed approach consistently outperforms the baselines in all settings.
    }
    \label{tab:benchmark}
\end{table*}

\begin{figure*}[!th]
    \centering
    \begin{subfigure}{.97\textwidth}
        \begin{tabular}{cc}
             \small Learnable linear layer  & \small Frozen linear layer \\
          \includegraphics[width=.47\linewidth]{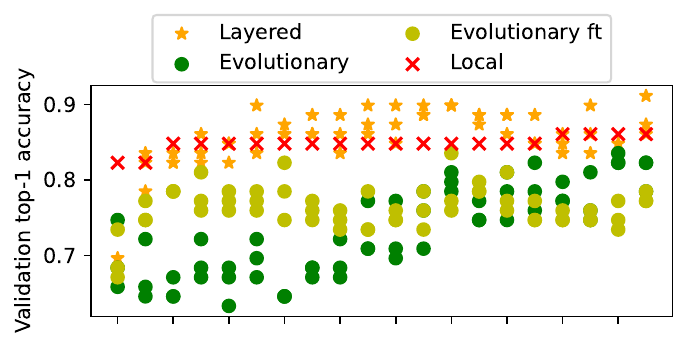}  & \includegraphics[width=.47\linewidth]{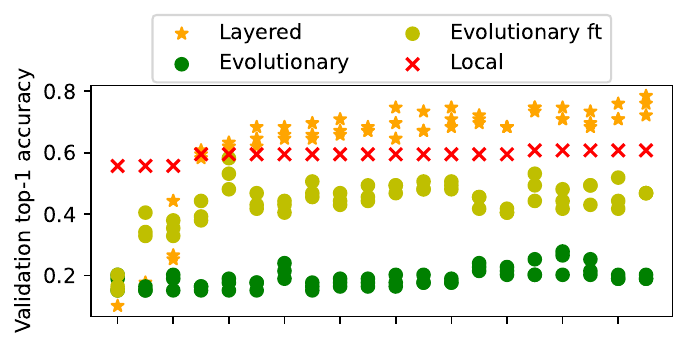} \\
        \end{tabular}
    \end{subfigure}
    \begin{subfigure}{.97\textwidth}
        \begin{tabular}{cc}
          \includegraphics[width=.47\linewidth]{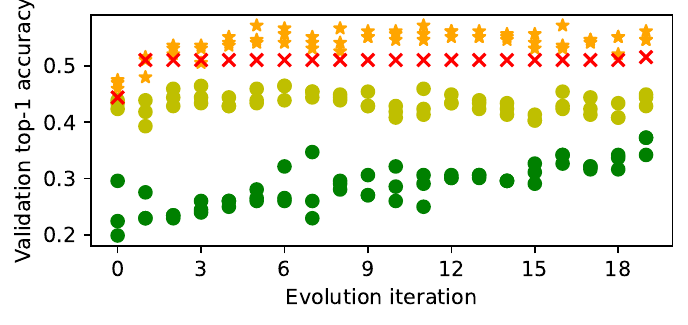}  &  \includegraphics[width=.47\linewidth]{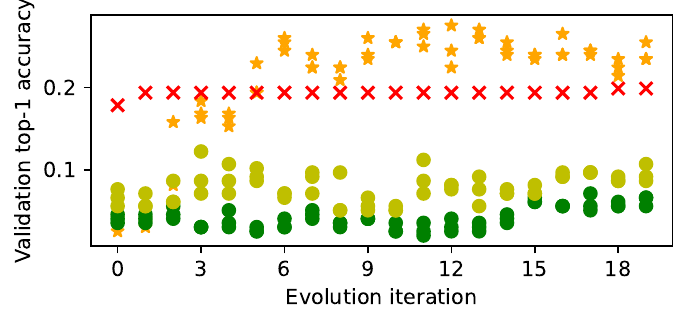}
        \end{tabular}
    \end{subfigure}
    \vspace{-12pt}
    \caption{\textbf{Convergence of the search algorithms on the reduced ModelNet10 (top) and ModelNet40 (bottom) datasets for models with learnable (left) and frozen (right) linear layers.}
    Shown are the accuracies of the top three candidates for the layered and evolutionary searches, and the single candidate for the local search.
    Layered search steadily improves accuracy by increasing PQC expressivity.
    Evolutionary search converges faster with fine‑tuning, while the version without fine‑tuning can occasionally find better models but struggles when the linear layer is frozen.
    Local search exhibits a step‑like behavior, as the model is updated only when an improvement is found.
    }
    \vspace{-3pt}
    \label{fig:convergence}
\end{figure*}

\subsection{Benchmark with QAS Methods}
\label{sec:qas_benchmark}
We ablate the proposed layered search against two alternatives: 1) An adaptation of a classical local search algorithm to the setting of PQCs, and 2) an evolutionary search approach similar to other QAS methods~\cite{du2022quantum,ma2024continuous}, making use of a super-circuit inspired by the classical one-shot architecture search approach in~\citet{guo2020single}. We evaluate the evolutionary search in two different variants: One where the parameters are directly taken from the super-circuit and one where the current architecture is fine-tuned. Evolutionary search models with fine-tuning are suffixed by ``ft''. Details as well as an algorithmic description for both approaches can be found in~\Cref{app:pseudo_codes}.

\begin{table*}[ht]
\centering
\resizebox{2.1\columnwidth}{!}{
\begin{tabular}{lcccccccccc}
                \hline
                \hline
                & \multicolumn{5}{l}{\textbf{ModelNet10}}& \multicolumn{5}{l}{\textbf{ModelNet40}} \\
                & \textsc{NPQ(L/F)}  & \textsc{NPC}   & \textsc{Tr.(L/F)}     & \textsc{Va.(L/F)}  & \textsc{Ts.(L/F)}
                & \textsc{NPQ(L/F)}  & \textsc{NPC}   & \textsc{Tr.(L/F)}     & \textsc{Va.(L/F)}  & \textsc{Ts.(L/F)}\\
                \hline
Layered$_{10}$ [Ours]          &   $100/207$    & $270$    &    $\mathbf{92/78 \%}$      &   $\underline{92}/\mathbf{78} \%$      &    $\underline{84}/\underline{67} \%$               &   $105/164$    & $1080$    &    $59/\mathbf{27} \%$       &   $56/\mathbf{26}\%$    &   $\underline{54}/\mathbf{21} \%$               \\
Layered$_{20}$ [Ours]          &   $279/282$    & $270$    &    $\mathbf{92/78 \%}$      &   $\mathbf{93/78 \%}$      &    $\mathbf{85/69 \%}$               &   $279/220$    & $1080$    &    $\mathbf{61}/\underline{22} \%$       &   $\mathbf{59}/\underline{22}\%$    &   $\mathbf{55}/\underline{16} \%$                        \\ 
Local search         &   $163/165$    & $270$    &    $\underline{90}/\underline{75} \%$      &   $89/\underline{75} \%$      &    $83/65 \%$               &   $166/166$    & $1080$    &    $\underline{60}/20 \%$      &   $\underline{58}/18 \%$      &    $54/13 \%$               \\ 
Evolutionary~\cite{guo2020single,ma2024continuous}          &   $162/163$    & $270$    &    $88/68 \%$       &   $85/65 \%$   &   $80/53 \%$              &   $162/168$    & $1080$    &    $59/18 \%$       &   $57/17\%$    &   $52/12 \%$              \\
Evolutionary ft.~\cite{guo2020single,ma2024continuous}      &   $163/176$    & $270$    &    $89/69 \%$       &   $85/68\%$    &   $81/56 \%$              &   $163/166$    & $1080$    &    $59/17 \%$       &   $58/16\%$    &   $53/11 \%$      \\       
 \hline   
\end{tabular}
    }
    \caption{\textbf{Ablation study on the choice of QAS:} \textsc{``NP(Q/C)''}=Number of parameters in the quantum or classical backbones; 
    \textsc{``Tr./Va./Ts.''}=Train/Validation/Test top-1 accuracy;
    \textsc{``L/F''}=Search models with learnable and frozen linear layers;
    {``ft.''}=Fine tuning.
    \textsc{NPC} is $0$ for models with frozen linear layers.
    The subscript in the layered model refers to the number of generations.
    }
    \label{tab:qas_benchmark}
\end{table*}
Our layered-QAS is performed over $20$ generations, which corresponds to $13$ parametrised layers and $7$ pruning layers added. 
In each generation, we train and evaluate $3$ candidate models.
The super-circuit for the evolutionary search also has $20$ layers, but the number of parameters per layer may be different than in the layered case. 
The super-circuit is trained with $100$ randomly sampled architectures.
The evolutionary search itself is performed over $20$ generations for a population of size $10$, from which only the top-$5$ architectures are considered to generate new ones.

\Cref{fig:convergence} presents the behaviours of the different search procedures on ModelNet10 and ModelNet40 over the course of the iterations, and in addition,~\Cref{tab:qas_benchmark} shows the training, validation and test accuracies for the best models found by each QAS for two different settings: Our standard setting in which the last (classical) linear layer is learnable as well as a setting with a frozen linear layer with random weights that gives an impression of how powerful the PQC alone (without classical components) is. 

The results of the evolutionary search are from the evolutionary search itself, \ie, after the super-circuit training.  
In~\Cref{app:performance}, we discuss the performance of the found models when trained from scratch after the search.

While all search strategies yield an increase in validation accuracy, one can see a clearly favourable behaviour of the layered search in both settings, with and without a learnable classical linear layer. 
It is interesting to observe the high jump of the layered-search accuracy of models with frozen linear layers at generation 2. 
This jump corresponds to the addition of the first entangled layer, which significantly increases the circuit's expressivity.

While the performance for all search strategies is higher in the learnable than in the frozen linear layer setting, it is remarkable that --at least in the case of ModelNet10-- our proposed approach reaches almost $80\%$ validation accuracy for a random linear layer, indicating a highly expressive PQC part. In comparison, we ran sQCNN-3D on ModelNet10 with a frozen MLP to obtain $17.4\%$ validation and $15.0\%$ test accuracy only, indicating a significant advantage in the expressiveness of our PQC architecture. 
In addition, our sQCNN‑3D-implementation yields lower accuracy than originally reported~\cite{baek2023stereoscopic}, likely because the original work employs a fully connected network (with no architectural details provided) for classification, whereas we employ only a single fully connected layer. 
This simplification is sufficient to highlight the superior accuracy–parameter trade‑off of our layered‑QAS approach.

\subsection{Ablation Studies and Runtimes}
\label{sec:ablation}
\begin{figure}[tb]
    \centering
    \includegraphics[width=.98\linewidth]{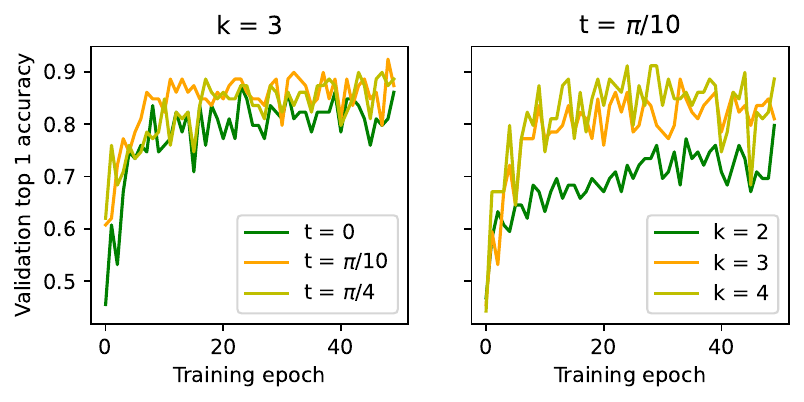}
    \caption{\textbf{Performance} of the layered search for different pruning thresholds $t$ and voxelisation granularity $k$ on the reduced ModelNet10.
    Pruning reduces the number of parameters without sacrificing accuracy.
    Increasing $k$ increases the prediction accuracy.
    }
    \vspace{-2pt}
    \label{fig:ablation}
\end{figure}

For the model with a learnable linear layer,~\Cref{fig:ablation} shows the convergence behaviour for different pruning thresholds $t$ and voxel granularity $k$ on the reduced ModelNet10.
We see that pruning gates with absolute parameter values below $t$ do not worsen the prediction accuracy while reducing the PQCs' depths. Pruning at $t=\pi/10$ and $t=\pi/4$ reduced the number of gates from $126$ to $86$ and $189$ to $148$ respectively. 
We note that other pruning methods based on Fisher information exist that guarantee not to affect the PQC expressivity~\cite{haug2021capacity}, but are more computationally expensive as they require the Fisher information matrix to be computed.

Finer voxel grids improve the prediction, as they allow capturing more details in the object shapes.
However, the performance for $k=4$ being only slightly better than for $k=3$ suggests that other factors are much more significant than the voxel granularity beyond $k=3$. 

\paragraph{Runtimes.} On the reduced ModelNet10, the layered search took  $2$ hours. 
Those times multiply by $3$ for the search on the reduced ModelNet40, and by $10$ on full datasets. 
Fine-tuning all search models took about $0.5$ and $1$ hours on the full ModelNet10 and ModelNet40 datasets, respectively. 
In comparison, the training of the sQCNN-3D and vanilla CNN took $10$ and $4$ hours on the full ModelNet10, which multiplies by $3$ on the full ModelNet40.

%% file: 5_summary.tex
\section{Discussion and Conclusion}
\label{sec:summary}

We evaluated PQCs discovered via our layered-QAS policy on 3D point cloud classification using the ModelNet datasets, combining amplitude encoding, QAS-found PQCs, and classical linear layers as principal components.
Our models outperformed the existing quantum sQCNN-3D and, when matched in expressivity to a purely classical baseline, achieved competitive accuracy with far fewer parameters.
Even with frozen classical layers, the QAS-discovered PQCs alone learned meaningful and discriminative features, underscoring the effectiveness of our architecture search.
These results demonstrate that layered-QAS can identify task-specific quantum architectures that balance expressivity and trainability, making them viable for challenging, structured domains like 3D classification.

A fundamental remaining problem is the under-expressivity of the considered models, mainly consisting of linear PQC operations, with quantum measurements being the only non-linear functions. 
Future work should investigate expressive PQC designs, such as those incorporating non-linearities via intermediate measurements.
Another interesting direction would be a binary encoding of the input in basis states, on which linear operations are provably expressive enough to approximate any output.
This alternative would require far more qubits, making it a goal for future quantum hardware \mbox{advancements}.

\paragraph{Acknowledgements.}
This work was supported by the Deutsche Forschungsgemeinschaft (DFG, German Research Foundation), project number 534951134.
NKM and MM acknowledge support by the Lamarr Institute for Machine Learning and Artificial Intelligence.

%% file: X_suppl.tex
\clearpage
\appendix
\maketitlesupplementary
\pagestyle{plain}

This supplementary material provides:
\begin{itemize}
    \item A short introduction to gate-based quantum computing necessary to understand the PQC-based model developed in the main paper,~\Cref{app:intro_qc}.
    \item A visualisation of the voxelised point clouds for different granularity $k$ on the modelNet10 dataset,~\Cref{app:voxelisation}. 
    \item Details pseudo-codes for the evolutionary and local search protocols,~\Cref{app:pseudo_codes}. 
    \item Details of the evolutionary and layered search strategies on the full ModelNet10 datasets, including results of the training from scratch of the found models,~\Cref{app:modelnet10_fullds}. 
    \item Confusion matrices of different benchmark models on the full test set of the ModelNet10 dataset,~\Cref{app:confusion_modelnet10}.
\end{itemize}

\section{Gate-based Quantum Computing}
\label{app:intro_qc}
We provide a short introduction to gate-based quantum computing necessary to understand our PQC models.
We introduce qubits, unitary transformations, and measurements, and give a brief discussion of challenges associated with the training of PQCs.

\paragraph{Qubit.}
The qubit is the fundamental unit of quantum information.
It is represented as a vector in the two-dimensional complex Hilbert space and is commonly written as:
\begin{equation}
\ket{\psi} = \alpha\ket{0} + \beta \ket{1},
\end{equation}
where $\alpha, \beta \in \mathbb{C}$ and are subject to the normalisation constraint $|\alpha|^2+|\beta|^2 = 1$.
This is because $\alpha$ and $\beta$ represent the probability amplitudes for physically measuring the qubit in the basis states $\ket{0}$ and $\ket{1}$, respectively.

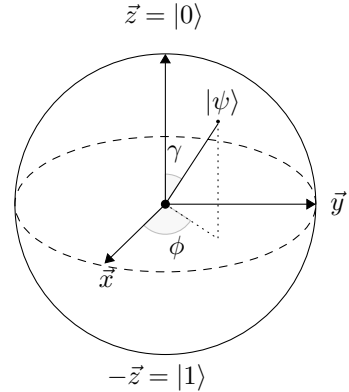
\begin{figure}[tbp]
    \centering
    \tikz \node[scale=1.0]{
    	\begin{tikzpicture}[line cap=round, line join=round, >=Triangle]
		\clip(-2.19,-2.7) rectangle (2.66,2.68);
		\draw [shift={(0,0)}, lightgray, fill, fill opacity=0.1] (0,0) -- (56.7:0.4) arc (56.7:90.:0.4) -- cycle;
		\draw [shift={(0,0)}, lightgray, fill, fill opacity=0.1] (0,0) -- (-135.7:0.4) arc (-135.7:-33.2:0.4) -- cycle;
		\draw(0,0) circle (2cm);
		\draw [rotate around={0.:(0.,0.)}, dash pattern=on 3pt off 3pt] (0,0) ellipse (2cm and 0.9cm);
		\draw (0,0)-- (0.70,1.07);
		\draw [->] (0,0) -- (0,2);
		\draw [->] (0,0) -- (-0.81,-0.79);
		\draw [->] (0,0) -- (2,0);
		\draw [dotted] (0.7,1)-- (0.7,-0.46);
		\draw [dotted] (0,0)-- (0.7,-0.46);
		\draw (-0.08,-0.3) node[anchor=north west] {$\phi$};
		\draw (-0.1,0.9) node[anchor=north west] {$\gamma$};
		\draw (-1.01,-0.72) node[anchor=north west] {$\vec{x}$};
		\draw (2.07,0.3) node[anchor=north west] {$\vec{y}$};
		\draw (-0.67,2.8) node[anchor=north west] {$\vec{z}=|0\rangle$};
		\draw (-0.9,-2) node[anchor=north west] {$-\vec{z}=|1\rangle$};
		\draw (0.4,1.65) node[anchor=north west] {$|\psi\rangle$};
		\scriptsize
		\draw [fill] (0,0) circle (1.5pt);
		\draw [fill] (0.7,1.1) circle (0.5pt);
	\end{tikzpicture}
    };

    \caption{Bloch sphere representation of a qubit.
    The qubit $\ket{\psi} = \alpha \ket{0} + \beta \ket{1}$ can be expressed as a unit vector in $\R^3$.
    Two angles $\gamma \in [0, \pi] $ and $\phi \in [0, 2\pi] $ fully describe the qubit in the basis spanned by the vectors $\vec{x}, \vec{y}$ and $\vec{z}$.}
    \label{fig:bloch_sphere}
\end{figure}

Due to this normalisation condition, the qubit can be rewritten as
\begin{equation}
\label{eq:blochqubit}
\ket{\psi} = e^{i\omega} \left(\cos\frac{\gamma}{2} \ket{0} + e^{i\phi} \sin \frac{\gamma}{2} \ket{1}\right),
\end{equation}
where the angles $\gamma \in [0, \pi]$ and $\omega, \phi \in [0, 2\pi]$.
This representation, called the \emph{Bloch-sphere} representation of $\ket{\psi}$, translates into a unit vector called the \emph{Bloch vector}
$\psi_{\textrm{bloch}} = (\cos\phi \sin\gamma, \sin\phi \sin\gamma, \cos \gamma)^\top \in \mathbb{R}^3$,
which can be visualised as a point on the three-dimensional unit sphere, as shown in~\Cref{fig:bloch_sphere}.

\paragraph{Composed Systems.}
A system composed of multiple separable qubits $\ket{\psi_1}, \ldots, \ket{\psi_k}$ has a state vector $\ket{\psi}$ that is expressed as the tensor product of the individual qubit state vectors:
\begin{equation}
\ket{\psi} = \ket{\psi_1} \otimes \ldots \otimes \ket{\psi_k}.
\end{equation}
These composed separable qubits span only a subset of the $2^k$-dimensional Hilbert space in which the computation takes place.
However, a quantum phenomenon known as entanglement allows multiple qubits to become correlated such that the resulting state vector $\ket{\psi}$ can no longer be expressed as a tensor product,
thus spanning the complete Hilbert space.
A popular example of such a state is the Bell state $\ket{\psi^+} = \frac{1}{\sqrt{2}} (\ket{00} + \ket{11})$.

\paragraph{Unitary Transformations.}
Quantum computation consists of transforming the initial state vector of the system into one that encodes the solution of a given problem.
In our point cloud classification case, this transformation creates a feature vector useful for classification.
Because the norm of the quantum state vector must always be $1$, the only valid transformations are unitary transformations, i.e., operators $\mathbf{U}$ satisfying
\begin{equation}
\mathbf{U} \mathbf{U}^\dagger = \mathbf{U}^\dagger \mathbf{U} = \mathbf{I}.
\end{equation}
Some common single-qubit operators are the Pauli $\textsc{X}, \textsc{Y}, \textsc{Z}$ operators, defined as:
\begin{equation}
\textsc{X} = \Matrix{ 0 & 1 \\ 1 & 0 }, \quad
\textsc{Y} = \Matrix{ 0 & -i \\ i & 0 }, \quad
\textsc{Z} = \Matrix{ 1 & 0 \\ 0 & -1 }.
\end{equation}
In learning tasks, since the exact state vector encoding the solution is unknown, the unitary transformation must be parametrised:
\begin{equation}
\label{eq:u3}
\mathbf{U} (\theta, \lambda, \phi) =
\Matrix{
\cos\left(\frac{\theta}{2}\right) & -e^{i\lambda}\sin\left(\frac{\theta}{2}\right) \\
e^{i\phi}\sin\left(\frac{\theta}{2}\right) & e^{i(\phi + \lambda)}\cos\left(\frac{\theta}{2}\right)
}.
\end{equation}
The $\textsc{RX}, \textsc{RY}, \textsc{RZ}$ gates in the main paper are special cases of this general operator.

For multi-qubit systems, the unitary operator can be written as the tensor product of single-qubit operators.
Entangled states can only be created through controlled operations, such as the controlled-NOT (CNOT) gate:
\begin{equation}
\textsc{CNOT} =
\Matrix{
1 & 0 & 0 & 0 \\
0 & 1 & 0 & 0 \\
0 & 0 & 0 & 1 \\
0 & 0 & 1 & 0
}.
\end{equation}
One can now parametrise such a controlled operation and let the optimisation decide whether to apply it or not.

\paragraph{Measurement.}
After the (Parametrised) unitary transformation, the system must be measured to extract useful information.
For a one-qubit system, the $\ket{0}$ and $\ket{1}$ basis states are eigenvectors of the Pauli-$\textsc{Z}$ observable, and measurement projects the state onto the $\textsc{Z}$ axis, reading out the $z$-coordinate:
\begin{equation}
\label{eq:measurement}
\braket{\textsc{Z}} = \braket{\psi|\textsc{Z}|\psi}.
\end{equation}
Since point cloud classification involves rotations around the $x, y, z$ axes of the Bloch sphere, measuring only in the $\textsc{Z}$ basis is insufficient; hence, all three Pauli bases are measured.

For multi-qubit systems, measurements can be performed globally, by tensoring the single-qubit measurement observables, or locally, by measuring only a subset of qubits.
It is important to mention that the measurement process is the only operation on state vectors that is not unitary and cannot be reversed.

\paragraph{Training PQCs.}
Training PQCs involves optimising circuit parameters to transform the initial state vector into a solution state.
Parametrised gates, such as in~\Cref{eq:u3}, are differentiable, and measurement, being a vector-matrix-vector multiplication as expressed in~\Cref{eq:measurement}, is also differentiable.
Classically simulated small-scale PQCs in PennyLane use backpropagation for optimisation. 
For large-scale PQCs, the parameter-shift rule enables computing gradients by evaluating the objective function twice per parameter~\cite{mitarai2018quantum}.

Training PQCs faces a major challenge known as the barren plateau, characterised by a flat loss landscape where an exponential number of measurements is needed to approximate gradients accurately~\cite{mcclean2018barren}. 
This arises from random states in high-dimensional Hilbert spaces, making measurements vanish on average. 
The Barren plateau provably has multiple causes ranging from data encoding, PQC architecture, measurement, and noise~\cite{larocca2024review}.
Techniques to mitigate this issue include local measurements, proper initialisation, and reducing PQC depth and width~\cite{larocca2024review}.

\section{Voxelisation}
\label{app:voxelisation}
\begin{figure}[ht]
    \centering
    \includegraphics[width=0.8\linewidth]{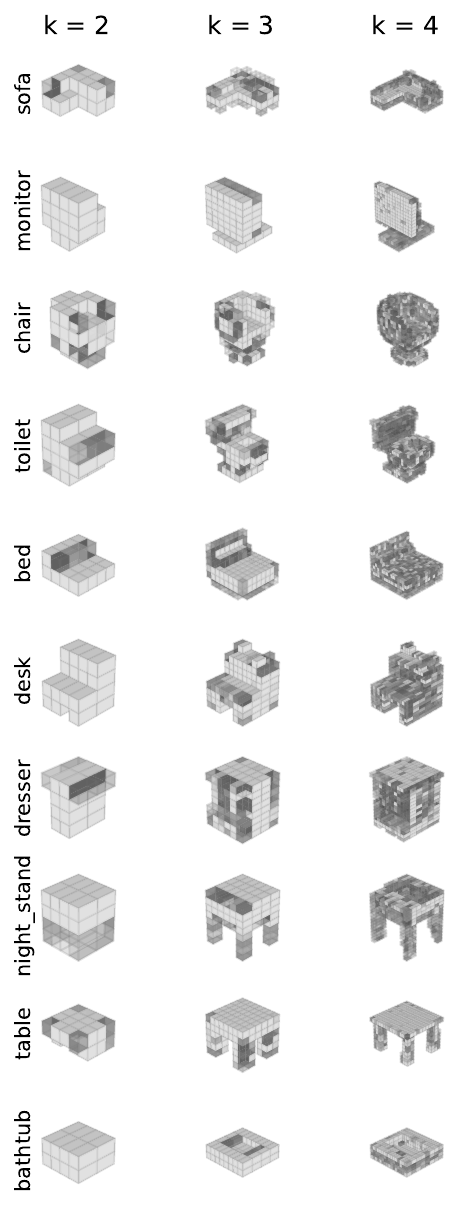}
    \caption{
    Voxelised point clouds with different granularity $k$.
    Brighter voxels are denser voxels, \ie voxels containing most of the points, and darker voxels are less dense voxels.
    Higher values of $k$ allow for capturing more details on the shape to classify.
    We use $k=3$ in our experiments.
    }
    \label{fig:visualisation}
\end{figure}

We provide exemplary visualisations of voxelised point clouds for different granularity $k$ in~\Cref{fig:visualisation} for the data labels \texttt{chair}, \texttt{sofa}, \texttt{table} from the ModelNet10 dataset.
We see that the density representation effectively captures finer details about denser and less dense regions of the shape, allowing for further understanding and classifying the shapes.
Without density information, the \texttt{chair} and the \texttt{toilet} at $k=3$ would more or less have the same shape, up to a rotation.
Larger values of $k$ further provide details about the data, but would require more resources and compute time to process the data.

\section{Evolutionary Search and Local Search Approaches for PQCs}
\label{app:pseudo_codes}
We provide more details, as well as pseudo-codes for the evolutionary and local search approaches that we use in \Cref{sec:qas_benchmark} of the main text.

\subsection{Evolutionary Search for PQC}
\label{app:evolutionary}
The evolutionary search approach is similar to other QAS methods~\cite{du2022quantum,ma2024continuous}. In our setting, we additionally make use of a super-circuit inspired by the classical one-shot architecture search approach in~\citet{guo2020single}. The super-circuit defines a useful search pool using shared parameters. 

Candidate architectures are then sampled randomly from the pool and trained for a few epochs to optimize the super-circuit parameters.

Once the super-circuit is trained, it undergoes an evolutionary search in which a population of architectures evolves to better explore the search space.

The evolutionary search algorithm is described in~\Cref{alg:algorithm_evolutionary}. As described above, 
it consists of two main phases: the training of the super-circuit and the evolutionary search itself.

\begin{algorithm}[t]
\caption{Evolutionary Quantum Architecture Search}\label{alg:algorithm_evolutionary}
\begin{algorithmic}[1]

\Require SuperCircuit, training and validation sets, population size, ranking metric $\text{Best}$.

\For{$i = 0, 1, 2, \ldots$} 
    \State \text{Sample} $\mat U_{\text{candidate}}$ from the SuperCircuit pool
    \State $\text{Train} \ \mat U_{\text{candidate}} \ \text{for a few epochs}$
\EndFor
 
\State \text{Set} $ m := \text{PopSise} / 2$
\State \text{Set} $ n := \text{PopSise} / 2$
\State \text{Initialise population} $ P_{0} := \text{Init} (\text{PopSise})$
\For{$i = 0, 1, 2, \ldots$} 
    \State $\text{Accuracies} = \text{(FineTuning+)Inference}(P_i, \text{ValSet})$
    \State \text{Update} $ \text{Topk} = \text{Best} (P_i, \text{Accuracies}, k)$
    \Comment{Ranking}
    \State \text{Set} $P_{\text{crossover}} = \text{Crossover}(\text{Topk}, m)$
    \State \text{Set} $P_{\text{mutation}} = \text{Mutation}(\text{Topk}, n)$
    \State \text{Update population} $P_{i+1} = P_{\text{crossover}} \cup P_{\text{mutation}}$
\EndFor
    \State \textbf{Return} $ \mat U = \text{Best} (P_i, \text{Accuracies}, 1)$
    \Comment{Ranking}
\end{algorithmic}
\end{algorithm}

\begin{figure}[t]
    \centering
	\includegraphics[scale=1., trim={0cm .1cm 0cm .2cm}, clip]{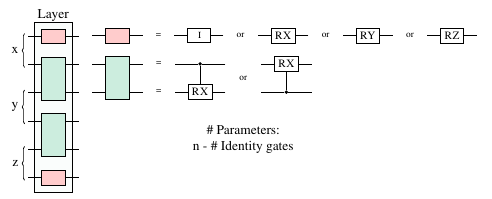}
    \caption{Candidate layers used in the evolutionary search. 
    Note that the gate types and qubits, including controlled and target qubits of controlled gates, are chosen randomly in each layer. 
    }
    \label{fig:candidate_layers_evolutionary}
\end{figure}

\paragraph{Search Pool.}
The super-circuit with shared parameters offers efficient storage management. 
For our application, the search space consists of several layers of gates, each involving a parametrised gate applied to each qubit.
A layer illustration is provided in~\Cref{fig:candidate_layers_evolutionary}.
Candidate gates for each qubit include the identity gate $\textsc{I}$ and Pauli $\RX, \RY, \RZ$ gates as single-qubit operations, as well as controlled $\CRX$ rotations for two-qubit interactions. 
For each qubit, any other qubit can be the target of the $\CRX$.
Thus, a layer consists of an application of one gate (single-qubit or controlled-gate) to each qubit.

\paragraph{Search Step.}
During the search itself, the architectures are ranked based on their performance on the validation set and optionally other user-supplied metrics.
Optionally, a fine-tuning of the models can be done before the inference.
The best-performing architectures are selected to be parents of the new generation. 
This new generation of architectures is generated through mutation and crossover operations applied to their parents and becomes the current population:
\begin{itemize}
    \item \textbf{Crossover}
    combines parts of two parent architectures to create a new architecture.
    The new architecture inherits gate sequences from either one parent or the other, consistently across all layers for each qubit.
    \item \textbf{Mutation}
    randomly alters some parts of the architecture.
    To do this, we randomly select a qubit and layer and change the corresponding gate.
\end{itemize}
The process is repeated until a user-defined number of generations is reached.

\subsection{Local Search for PQC}
\label{app:local}
We also adapt the classical local search~\cite{white2021exploring} to the setting of PQCs by making use of the trained super-circuit from the evolutionary search approach. 
As this super circuit was already trained on a search pool for the evolutionary search, we leverage the trained parameters of the super-circuit to optionally just fine-tune the candidate models for one epoch before inference on the validation set, instead of training all candidate architectures in the neighbourhood of the current model from scratch as described in~\citet{white2021exploring},

We start from a random architecture, randomly select a qubit and layer, change the corresponding gate, and accept the change if it improves the QPC performance on the validation set after one fine-tuning epoch.

We provide the detailed algorithm used for the local search in~\Cref{alg:algorithm_local}.
\begin{algorithm}[t]
\caption{Local Quantum Architecture Search}\label{alg:algorithm_local}
\begin{algorithmic}[1]

\Require SuperCircuit, training and validation sets, Neighbourhood function, ranking metric $\text{Best}$.

\State \text{Sample} $\mat U$ from the SuperCircuit pool

\For{$i = 0, 1, 2, \ldots$} 
    \State $\text{ArchList} := [\ ]$
    \Comment{Empty list}
    
    \For{$k = 0, 1, 2, \ldots, \text{NumCandidates}$} 
        \State $\text{Pool}_k$ = SuperCircuit $\cap$ Neighborhoud$(\mat U)$
        \State \text{Sample} $\mat U_{\text{candidate}}$ from $\text{Pool}_k$
        \State Optionally fine-tune $\mat U_{\text{candidate}}$ for one epoch
        \State Append $\mat U_{\text{candidate}}$ to ArchList
    \EndFor
    \State \text{Update} $ \mat U = \text{Best}(\text{ArchList})$
    \Comment{Ranking}
\EndFor
    \State \textbf{Return} $ \mat U$
\end{algorithmic}
\end{algorithm}

\section{Search on the Full ModelNet10 Dataset}
\label{app:modelnet10_fullds}
We describe the behaviours of the layered and evolutionary models when trained on the full, large ModelNet10 dataset.

\subsection{Analysis of the Search Behaviours}
\label{app:convergence}

As mentioned in~\Cref{sec:results}, we observed that the search algorithms, principally the evolutionary search, struggle to identify better models when using the full datasets, in addition to increasing the training time.
The principal reason is that $5$ epochs of training candidate models on the full, large dataset is sufficient for the model parameters to converge.
As a consequence, the weight-sharing mechanism of the evolutionary search leads to conflictual parameter updates.

\begin{figure}[t]
    \centering
        \begin{subfigure}{.48\textwidth}
            \includegraphics[width=.98\linewidth]{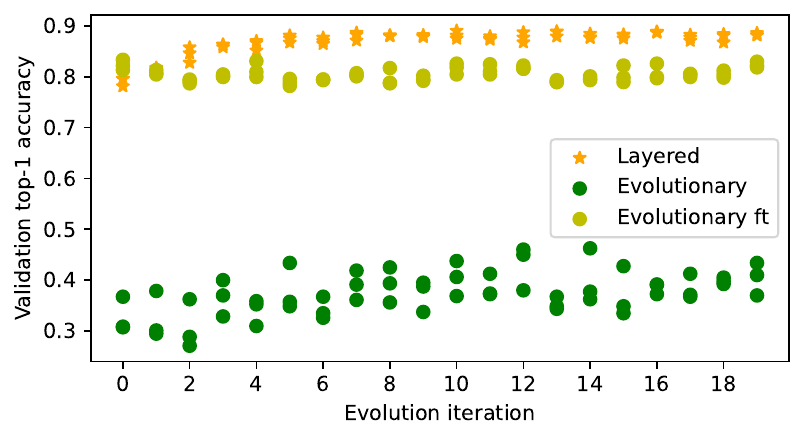}
            \caption{Learnable linear layer}
        \end{subfigure}
        \hfill
        \begin{subfigure}{.48\textwidth}
            \includegraphics[width=.98\linewidth]{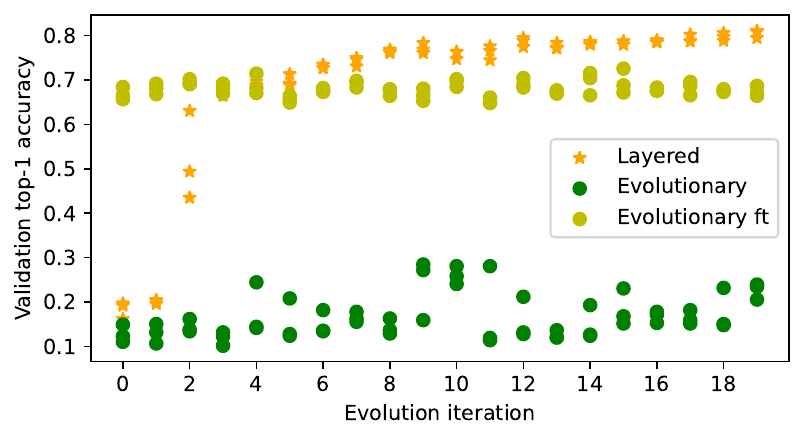}
            \vspace{-.2cm}
            \caption{Frozen linear layer}
        \end{subfigure}
    \caption{Convergence plot of the search algorithms on the full ModelNet10 dataset.
    Shown are the accuracies of the three best candidate models for each search strategy.
    The Layered search improves the prediction accuracy by enhancing the PQC expressivity.
    The evolutionary search, even with fine-tuning, shows only slight improvement over the generations.
    }
    \label{fig:convergence_m10}
\end{figure}

\Cref{fig:convergence_m10} presents the convergence behaviour of the search algorithms on the full ModelNet10 dataset, with the same training configuration as in the main paper. 
We see that both the evolutionary search variants, with and without fine-tuning, face difficulties in identifying good models over generations. 
In contrast, the layered search continues to improve prediction accuracy, with competition between candidate models becoming tighter as generations progress. 
The interpretation is that sufficient training data underscores the strengths of the models, leading to improved overall performance.

\subsection{Performance Evaluation}
\label{app:performance}

We next report performance results of the best models found on the full ModelNet10 dataset.
The best model of each search strategy is the model with the highest validation top-1 accuracy after search.
At the same time, we investigate whether or not the warm start of the candidate models is beneficial in optimising the found architectures.
To this end, we: (i) Fine-tune the found models over $10$ more epochs with the reduced learning rate (we call this fine-tuning after search); and (ii) Train the found models from scratch with randomly initialised parameters for $20$ epochs, followed by fine-tuning for $10$ more epochs (we call this training from scratch).
The parameter range for random initialisation is $[-10^{-2}, 10^{-2}]$.

\begin{figure}[t]
    \centering
        \begin{subfigure}{.48\textwidth}
            \includegraphics[width=.48\linewidth]{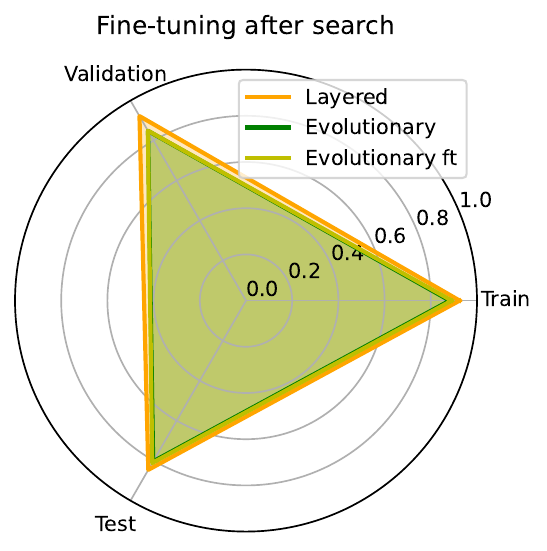}
            \includegraphics[width=.48\linewidth]{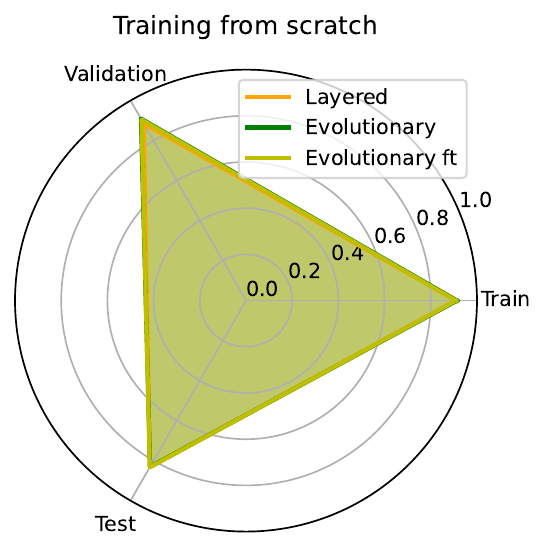}
            \caption{Learnable linear layer}
        \end{subfigure}
        \begin{subfigure}{.48\textwidth}
        \vspace{.2cm}
            \includegraphics[width=.48\linewidth]{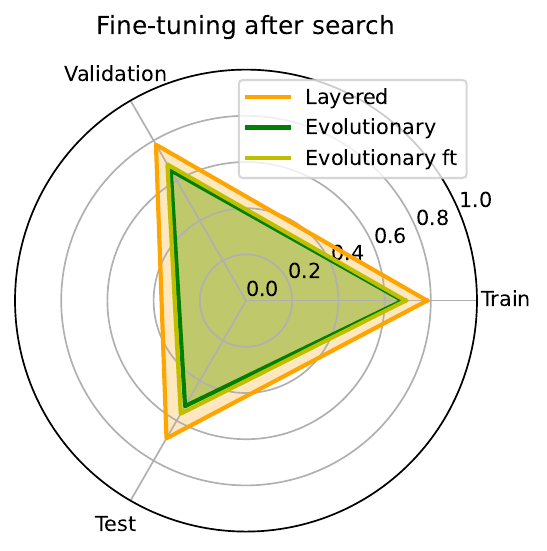}
            \includegraphics[width=.48\linewidth]{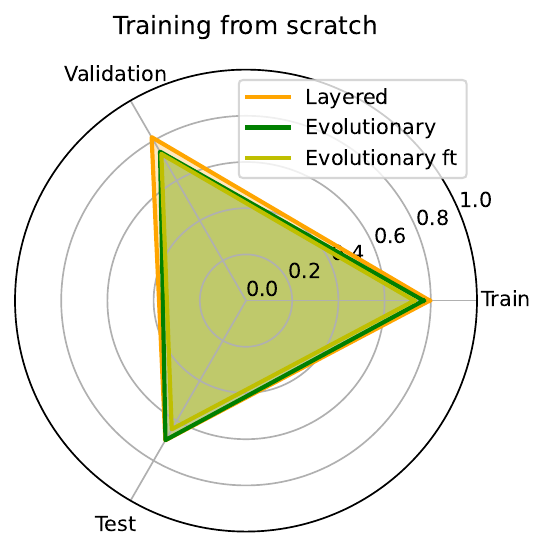}
            \caption{Frozen linear layer}
        \end{subfigure}
    \caption{Benchmark performance of found models.
    Reported are top-1 accuracies on the respective sets.
    }
    \label{fig:performance_m10}
\end{figure}

\Cref{fig:performance_m10} showcases the performance of the top models identified by the layered and evolutionary search methods on the full ModelNet10 dataset. 
The layered model significantly outperforms the evolutionary models, with the fine-tuned evolutionary model offering only a slight improvement over the non-fine-tuned version. 
There is a substantial discrepancy between the accuracies of the evolutionary models after fine-tuning and those trained from scratch, with the latter performing better. 
In contrast, the layered model trained from scratch achieves a similar accuracy to that obtained after fine-tuning. 
This indicates that the validation accuracies relied upon by the evolutionary search are highly questionable.

Indeed, search methods using super-circuits or supernetworks, as phrased in common machine learning approaches, are based on the assumption that the performance of models using the weight-sharing weights is correlated with the performance of the models being trained from scratch. However, several works discuss the validity of this assumption, arguing that the correlation is heavily dependent on the defined search space and the training of the supernetworks~\cite{Yu2020NAS, Zhang2020NAS}. Thus, the results of the models using weight-sharing weights and training them from scratch can lead to different performances.

\newpage

\subsection{Confusion Matrices on ModelNet10}
\label{app:confusion_modelnet10}

\begin{figure*}[!ht]
    \centering
    \hspace{1cm}
        \begin{subfigure}{.45\textwidth}
            \includegraphics[width=.8\linewidth, trim={0cm 0cm 0cm 0cm}, clip]{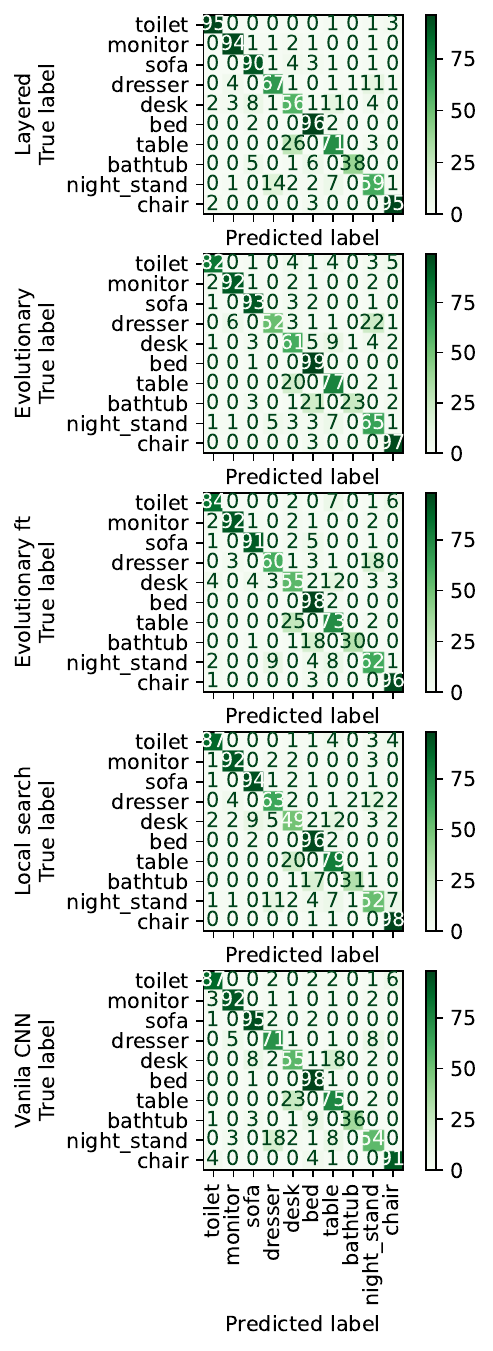}
            \caption{Learnable linear layer}
        \end{subfigure}
        \begin{subfigure}{.45\textwidth}
            \includegraphics[width=.8\linewidth, trim={0cm 0cm 0cm 0cm}, clip]{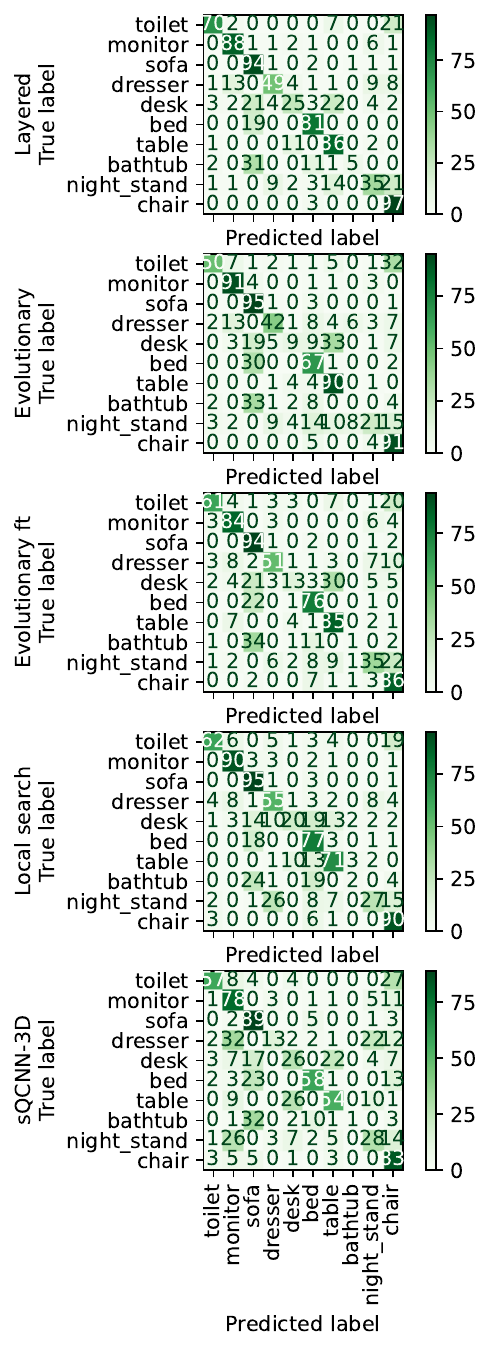}
            \caption{Frozen linear layer}
        \end{subfigure}
    \caption{Confusion matrices on the test set of the ModelNet10 dataset.
    Models with learnable linear layers are more accurate than those with frozen linear layers.
    All models consistently misclassify similar shapes like \texttt{desk}\&\texttt{table}, \texttt{bathtub}\&\texttt{bed}, or \texttt{toilet}\&\texttt{chair}.
    The latter pair is particularly difficult for models with frozen linear layers.
    Note that learnable and frozen linear layers do not apply to the vanilla CNN and sQCNN-3D models on the last row.
    ``ft.'' stands for fine tuning.}
    \label{fig:confusion_m10}
\end{figure*}

We now take a closer look at the classification challenges on ModelNet10.
Confusion matrices are provided in~\Cref{fig:confusion_m10} for all the benchmark models.
We observe that some class pairs are more challenging to classify than others.
For instance, all the models show difficulties in distinguishing \texttt{desk} \& \texttt{table} or \texttt{toilet} \& \texttt{chair}, since those shapes visually also look very similar.
The latter pair is particularly difficult for models with frozen linear layers.
The misclassification may be attributed to the coarse voxelisation granularity and the limited expressivity of the models.
From the confusion matrices, however, we can see again that the layered model, both with learnable and frozen linear layers, can better classify challenging pairs than other models.

